\documentclass[english,aps,superscriptaddress,showkeys,showpacs,prepri]{revtex4}
\usepackage[T1]{fontenc}                                                        
\pdfoutput=1                                                                    
\usepackage[letterpaper]{geometry}                                              
\usepackage{wrapfig,rotating}                                                   
\usepackage{units}                                                              
\usepackage{bbding}                                                             
\usepackage{amsmath}                                                            
\usepackage{amssymb}                                                            
\usepackage{graphicx}                                                           
\usepackage{esint}                                                              
\usepackage[colorinlistoftodos,prependcaption,textsize=tiny]{todonotes}
\usepackage{listings}
\usepackage{pbox}
\usepackage{tabularx}

\usepackage[utf8]{inputenc}                                                     
\usepackage{lmodern} 
 
\makeatletter

\DeclareGraphicsExtensions{.png}
\graphicspath{ {../},{./} }

\usepackage{doi}                                                                
\usepackage{hyperref}                                                           
                                                                                
\makeatother                                                                    
                                                                                
\usepackage{babel}

\begin{document}

\title{GPU acceleration and performance of the particle-beam-dynamics code Elegant}

\author{J.R. King}
\affiliation{Tech-X Corporation, Boulder CO 80303, USA}

\author{I.V. Pogorelov}
\altaffiliation{Presently at RadiaSoft LLC}
\affiliation{Tech-X Corporation, Boulder CO 80303, USA}

\author{K.M. Amyx}
\altaffiliation{Presently at Sierra Nevada Corporation}
\affiliation{Tech-X Corporation, Boulder CO 80303, USA}

\author{M.~Borland}
\affiliation{Argonne National Laboratory, Argonne, IL 60439, USA}

\author{R.~Soliday}
\affiliation{Argonne National Laboratory, Argonne, IL 60439, USA}

\date{\today}
\begin{abstract}
Elegant is an accelerator physics and particle-beam dynamics code widely used
for modeling and design of a variety of high-energy particle accelerators and
accelerator-based systems.  In this paper we discuss a recently developed
version of the code that can take advantage of  CUDA-enabled graphics
processing units (GPUs) to achieve significantly improved performance for a
large class of simulations that are important in practice.  The GPU version is
largely defined by a framework that simplifies implementations of the
fundamental kernel types that are used by Elegant: particle operations,
reductions, particle loss, histograms, array convolutions and random number
generation.  Accelerated performance on the Titan Cray XK-7 supercomputer is
approximately 6-10 times better with the GPU than all the CPU cores associated
with the same node count. In addition to performance, the maintainability of
the GPU-accelerated version of the code was considered a key design objective.
Accuracy with respect to the CPU implementation is also a core consideration.
Four different methods are used to ensure that the accelerated code faithfully
reproduces the CPU results.
\end{abstract}



\maketitle


{\bf PROGRAM SUMMARY}

\begin{small}
\noindent
{\em Program Title:} Kernels from the GPU-accelerated Elegant  \\
{\em Licensing provisions: MIT} \\
{\em Programming language:} C/C++/CUDA \\
{\em Nature of problem:} 
The original design of the Elegant accelerator physics code was implemented on 
central processing units with message-passing interface parallelization. 
This implementation is not able to use next-generation multicore systems. \\
{\em Solution method:}
In this package we develop routines based on the CUDA language extensions to
C++ that enable porting the Elegant code to be run on graphics processing units
(GPUs). Special consideration is given to algorithms that require collective
communication on the GPU. \\
{\em Additional comments including Restrictions and Unusual features:}
The full Elegant source code is freely available from Argonne National
Laboratory and these distributions include the GPU code in the later releases. \\

\end{small}

\section{Introduction}
\label{sec:introduction}
{\parindent 16pt

Elegant is an open-source, multi-platform code used for design, simulation, and
optimization of a wide variety of high-energy particle accelerators and
accelerator-based systems, including free-electron laser (FEL) driver linear
accelerators (``linacs''), energy recovery linacs (ERLs), and storage rings
\cite{borland00, borland09,eleganthome}. The parallel version, Pelegant
\cite{wang07,shang09,wang09}, uses MPI for parallelization and shares all source
code with the serial version.  In a number of settings that include
accelerator design optimization, Elegant is used as the tracking component of
fully scripted simulations.  Elegant is fundamentally a lumped-element
particle accelerator tracking code utilizing 6D phase space, and is written
mostly in C. A variety of numerical techniques are used for particle
propagation, including transport matrices (up to third order), symplectic
integration, and adaptive numerical integration. Collective effects are also
available, including space charge, coherent synchrotron radiation (CSR),
wakefields, and resonant impedances.

In recent years, general purpose computing on graphics processing units (GPUs)
has attracted significant interest from the scientific computing community
because these devices offer unmatched performance at low cost and at high
performance per watt. Unlike general purpose processors, which devote
significant on-chip resources to command and control, pre-fetching, caching,
instruction-level parallelism, and instruction cache parallelism, GPUs devote a
much larger amount of silicon to maximizing memory bandwidth and raw
floating-point computation power. This comes at the expense of shifting the
burden towards developers and away from on-chip command and control logic,
and additionally requires relatively large problems with high levels of
parallelism.

One of the challenges of accelerating a code such as Elegant is the shear
number and variety of kernels required to accelerate common use cases. Without
reasonable accelerated coverage of the code the benefits of using the GPU may
be severely reduced. This reduction occurs both from the time required to
transfer the particles between the device and host memory when entering a stage
of a simulation that cannot be performed on the GPU, as well as due to the
fundamental limit in the form of  Amdahl's argument \cite{amdahl67}. Amdahl's
argument states that if a runtime fraction, $F$, of a code is accelerated
(threaded) with $n$ concurrent threads, then the maximum speedup is
$(F+(1-F)/n)^{-1}$. Thus, the speedup from an accelerated portion of a code that
covers a runtime fraction of 50\% with infinite threads is only a factor of
two, a rather modest acceleration. It has been our intent to accelerate a
sufficient number of elements such that most beamline computations contain a
significantly large runtime fraction of accelerated code.  In addition, we
intend the framework through which this acceleration is implemented to be
extensible such that new beamline elements may be added as necessary.

The paper is organized as follows.  Section \ref{sec:infrastructure} describes
the acceleration framework.  The vast majority of accelerated elements use a
C++ templated class framework to greatly simplify implementation of
per-particle operations. Depending on the element, the accelerated code may
also use the framework routines to form histograms, perform array convolutions,
evaluate reduction operations, and account for particle losses. In
Section~\ref{sec:performance} the performance of the accelerated functions is
presented. We compare a NVIDIA Tesla K20c and Volta V100 GPUs with a single-core of a Intel
Core i7-3770K CPU. This is inherently an unfair comparison as a workstation
typically contains 4 to 16 cores. However, the speedup factors are not the
focus of this section, rather we intend it as a discussion of the challenges of
accelerating various functions and the scaling of the accelerated code with the
number of particles. A fair comparison of full application speed-up is
presented in Section~\ref{sec:scaling} for a realistic particle-accelerator
simulation use case. Our implementation maintains Elegant's existing MPI
infrastructure to support CUDA-MPI hybrid parallelism. In this section we
compare performance of the application on the Titan Cray XK-7 on both the CPUs
and GPUs. Finally, in Section~\ref{sec:verification} we describe the multiple
verification techniques used to ensure the GPU-accelerated code faithfully
reproduces the original CPU-only implementation results.

}
\section{Computational infrastructure}
\label{sec:infrastructure}
{\parindent 16pt

The accelerated version of Elegant is programmed with the CUDA programming
model; CUDA is NVIDIA's programming model for GPU computing \cite{cuda,cuda08}.  In
general, the organization of the accelerated Elegant code attempts to mirror
the file organization of the CPU code as much as practical. For example, the
code for the \textit{csbend}, \textit{csr\_csbend} and \textit{drift\_csbend}
elements~\cite{ElegantManual} is located in the file csbend.c and the
accelerated kernels used by these elements are located in the file
gpu\_csbend.cu (the file suffixes cu and hcu are used for files containing
CUDA code). 

Generally speaking, a particle tracking simulation with Elegant consists of the
particle beam traversing a succession of so-called elements, each element
representing a magnetic optics component, a physical effect, or a combination
of the two~\cite{ElegantManual}.  Currently, not all elements have
GPU-accelerated implementations; we have focused our efforts on the most
commonly used elements. However, any beamline will run with the GPU-accelerated
version of Elegant. Silent support is provided where the particles are
transferred between the GPU and CPU if a beamline contains an element kernel
which has only a CPU implementation.

We next describe aspects the GPU framework used to simplify the process of
accelerating elements. There are three major barriers to acceleration: (1) the
CPU code uses an array-of-structs data format whereas the GPU code must use a
struct-of-arrays data format which permits coalesced memory transactions, (2)
elements with collective effects require reduction, convolution and histogram
operations, and (3) a sort operation is required in elements where particles
are lost. 


\subsection{Template meta-programming} \label{sec:TMP}

Compile-time polymorphism allows abstraction layers that can hide the more
close-to-the-metal implementation details from application developers. We
exploit compile-time polymorphism through template meta-programming with the
following aims:
\begin{itemize}\addtolength{\itemsep}{-0.5\baselineskip}
\item{To create extendible kernels that reduce both code maintenance and
programming errors.}
\item{To provide abstract interfaces that hide the CUDA-specific data-parallel
implementation details.}
\item{To ease the development workflow by avoiding CUDA-related boilerplate
such as thread and block configurations, thread-index computations, and
conversion between array-of-structures and structure-of-arrays data
formats.}
\end{itemize}

Central to this model is the GPU particle accessor class that behaves as though
it were a CPU-style array of structs. Given such an accessor, a developer need
only define a functor that acts on the individual particles by creating a basic
class and overloading the proper operator. Any non-particle data needed by the
functor, such as physical constants and auxiliary arrays, is placed in the
class member variables. The developer then passes the class to a template GPU
driver function and thus replaces a complicated for-loop over particles with a
small functor class and a call to a GPU driver. Minimal explicit CUDA code is
written; instead, a developer writes per-particle update classes that
encapsulate the data needed and algorithm to perform each particle update step.

\begin{lstlisting}[frame=single, language=c++, 
  caption={The Elegant CPU exactDrift function.},
  label={code:cpuExactDrift}]
#define sqr(x) (x*x)
void exactDrift(double **part, long np, double length)
{
  long i;
  double *coord;
  for (i=0; i<np; i++) {
    coord = part[i];
    coord[0] += coord[1]*length;
    coord[2] += coord[3]*length;
    coord[4] += length*sqrt(1+sqr(coord[1])+sqr(coord[3]));
  }
}
\end{lstlisting}

\begin{lstlisting}[frame=single, language=c++, 
  caption={The Elegant GPU exactDrift function.},
  label={code:gpuExactDrift}]
class gpuExactDrift{
public:
  gpuExactDrift(double len) : length(len) {}
  __device__ void inline operator()(gpuParticleAccessor& coord){
    coord[0] += coord[1]*length;
    coord[2] += coord[3]*length;
    coord[4] += 
      length*sqrt(1+coord[1]*coord[1]+coord[3]*coord[3]);
  }
  double length;
};
void gpu_exactDrift(long np, double length){
  gpuDriver(np, gpuExactDrift(length));
}
\end{lstlisting}

For illustration, we consider the Elegant \textit{exactDrift} kernel which has
a CPU implementation as given in Code~\ref{code:cpuExactDrift}. Note the GPU
implementation, Code~\ref{code:gpuExactDrift}, only contains a single CUDA
keyword, the \_\_device\_\_ keyword which instructs the compiler to compile the
function operator() for the GPU. Even though the particle data is stored in
struct-of-arrays format on the GPU, it is accessed through the GPU
particle-accessor-class bracket operator just like the corresponding CPU data.
The kernel as written is almost identical to that on the CPU although under the
hood vastly different data structures are used. The GPU particle accessor
class reads particle data from the struct-of-arrays format into the thread
registers only if they are used.  Upon destruction of the class (which is
associated with a specific particle) particle data associated only with the
indices used in the kernel are written.  This workflow leads to nearly optimal
global memory access patterns as the particle data can at most have a single
read/write operation per kernel where the results of intermediate operations
are stored within thread registers.

\subsection{Histogram operations} \label{sec:HO}

\begin{figure}
\centering
  \includegraphics[width=0.5\linewidth]{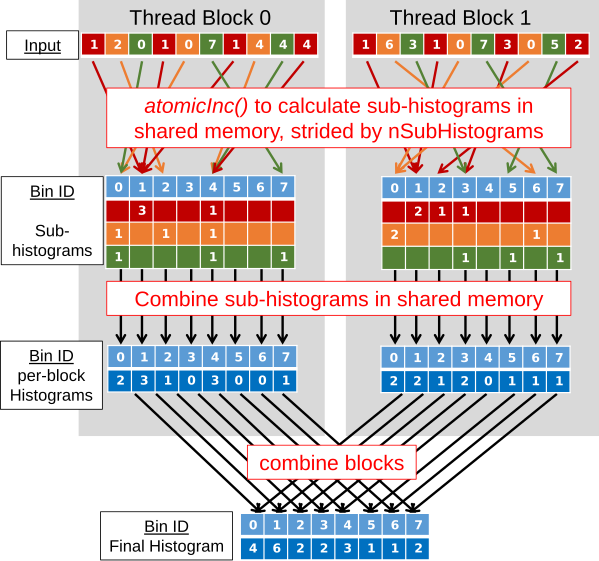}
  \caption{A pictorial representation of the GPU Elegant histogram algorithm.}
  \label{fig:histogram}
\end{figure}

Many of the beamline elements in Elegant that incorporate collective effects
utilize a histogram-based approach. Histogram calculations on a GPU are
challenging. In a straightforward GPU implementation, threads increment
histogram bins that are present in shared memory. However, as multiple
threads attempt to increment the same location in memory at the same time, this
leads to thread contention issues that may cause performance degradation (if
thread-safe atomic operations are used) or race conditions (otherwise). Atomic
operations to shared memory are not efficient, and the computational cost of
such atomics roughly scales with the level of thread contention. For histogram
kernels, the level of thread contention is largely determined by the number of
bins and the resulting distribution (for example, a distribution that is
localized to a few bins will create large thread contention). In order to
minimize the computational cost of these atomic operations, the GPU Elegant
histogram algorithm creates sub-histograms which reduce the thread contention
but require an additional operation to subsequently combine the
sub-histograms. This algorithm is illustrated pictorially in
Fig.~\ref{fig:histogram}. The GPU histogram kernel is optimized to fit as many
sub-histograms as possible per thread block while maintaining high block
occupancy. The number of thread blocks is limited by the optimal block
occupancy for a given device multiplied by the number of multiprocessors.

\subsection{Convolution operations}

Another potential barrier to achieving a good speedup on full beamline
computations is array convolution operations (such as those present in
wakefield elements).  A serial implementation of this algorithm scales as
$O(N_1\times N_2)$, where $N_1$ and $N_2$ are the sizes of the arrays in the
convolution. For typical operations, where arrays are histograms, the array
size is several hundred to a few thousand.  Given the convolution's 
$O(N^2)$ scaling, we cannot afford to rely on the serial CPU calculation without
negatively impacting performance of the simulation as a whole, even with
millions of particles.

Our GPU kernel achieves good acceleration by buffering sub-sections of each
array in shared memory while performing $O$(buffer size) computations. This
operation computes part of the final result for a given array index. As the
convolution is a linear operation, each thread block then applies an
atomic addition operation to produce the final result of the convolution.

\subsection{Reduction operations with asynchronous execution} \label{sec:ROAE}

Reductions of particle quantities to determine statistical beam properties are present in
many Elegant beamline elements. We template standard reduction algorithms over
the reduction operation (e.g.~sum, minimum, maximum, etc.). In these algorithms
thread blocks concurrently apply the reduction operation to subsections of the
data array and place the result in global memory.  The last block to finish
the sub-reduction then reduces the results from the previous step. 

Certain functions (i.e.~\textit{accumulate\_beam\_sums} and
\textit{compute\_centroids}) compute the beam properties and may be called from
multiple elements or the main Elegant \textit{do\_tracking} loop. These functions reduce
quantities from separate data arrays, for example during computations of the
mean and standard deviation of beam's position and/or transverse velocity. By
launching these functions asynchronously on separate CUDA streams we can take
advantage of concurrency during the last reduction operation and the transfer
of the reduction result(s), in addition to moving towards achieving the maximum
device memory bandwidth during the concurrent reductions. Asynchronous
reductions are 40\% faster than their synchronous counterparts with data arrays
of size one million on a NVIDIA Tesla K20c.

\begin{lstlisting}[frame=single, language=c++, 
  caption={Part of the Elegant CPU compute\_centroids code},
  label={code:cpuComputeCentroids}]
  for (i_part=0; i_part<n_part; i_part++) {
    part = coordinates[i_part];
    for (i_coord=0; i_coord<6; i_coord++)
#ifndef USE_KAHAN
      sum[i_coord] += part[i_coord];
#else
      sum[i_coord] 
        = KahanPlus(sum[i_coord], part[i_coord], &error[i_coord]);
#endif
  }
\end{lstlisting}

\begin{lstlisting}[frame=single, language=c++, 
  caption={Part of the Elegant GPU compute\_centroids code},
  label={code:gpuComputeCentroids}]
  for (i_coord=0; i_coord<6; i_coord++)
#ifndef USE_KAHAN
    gpuReduceAddAsync(d_particles+particlePitch*i_coord, 
                      n_part, &sum[i_coord]);
#else
    gpuKahanAsync(d_particles+particlePitch*i_coord, &sum[i_coord],
                  &error[i_coord], n_part);
#endif
  finishReductionStreams();
\end{lstlisting}

Our framework for these asynchronous reductions only requires the user to call
the function \textit{finishReductionStreams} before the result of the reduction
is used. For illustration, consider the CPU code shown in
Code~\ref{code:cpuComputeCentroids} and the GPU code in
Code~\ref{code:gpuComputeCentroids}. Aside from the different function calls,
and the removal of the loop over particles which is placed inside the GPU reduction 
calls, the only other difference between these two sections of code is the addition
of the \textit{finishReductionStreams} call to the GPU code. The management of the
CUDA streams, and allocation and use of pinned memory for the asynchronous memory transfer
between the device and the host is generalized within the GPU Elegant computational
framework.

\subsection{Particle losses and sorting} \label{sec:PLS}

Many beamline elements allow for particle losses (for example, a particle may
collide with an aperture edge). When a particle is lost on the CPU, it is
swapped with the particle at the end of the particle array and the particle
count is decremented. This algorithm is not amenable to the GPU which performs
concurrent particle-update operations. A straightforward GPU algorithm is to
fill an array with the particle index plus the number of particles if the
particle is lost, and just the particle index otherwise, and then sort the
particle array by this key. However, this too is somewhat inefficient as sort
algorithms are not amenable to the concurrency of the GPU. One can reasonably
expect that the fraction of lost particles for any given element is small
($<10\%$), otherwise the particular beamline configuration would not be of
great interest. We use this property to create a more efficient algorithm than
the straightforward sort-by-key algorithm. Relative to the sort-by-key
algorithm, this optimized algorithm is $4\times$ faster with $0.5\%$ losses,
$3\times$ faster with $5\%$ losses, $2.5\times$ faster with $10\%$ losses and
roughly equivalent with $50\%$ losses (benchmarking on an NVIDIA Tesla K20c).

\begin{figure}
\centering
  \includegraphics[width=0.5\linewidth]{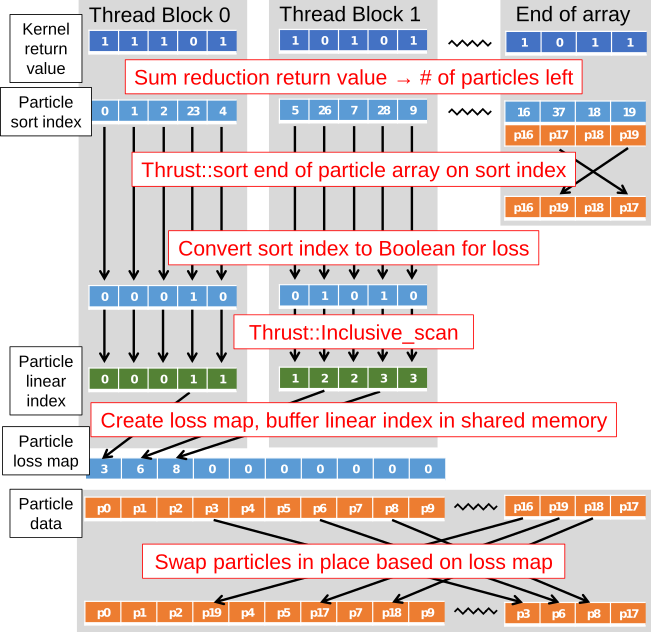}
  \caption{A pictorial representation of the GPU Elegant particle-loss sorting algorithm 
           for a case with 20 total and four lost particles.}
  \label{fig:killParticles}
\end{figure}

Our particle-loss algorithm is illustrated in Fig.~\ref{fig:killParticles}.  A
computational kernel must do two things to incorporate particle losses: It
should return an unsigned integer (zero if the particle is lost and unity
otherwise). It should also fill a particle-sort index with the particle index
plus the number of particles if the particle is lost, and the particle index
otherwise. The particle-loss algorithm does a sum reduction over the return
value. If the result is equal to the number of particles, no particles are lost
and the remainder of the loss computation is skipped. If particles are lost,
the end of particle array (size of the number of lost particles) is sorted, and
then sort index is converted to unity if the particle is lost, and zero
otherwise. An inclusive scan is performed which creates a particle linear index
array. When subsequent elements of this array are different, a particle is lost
and the value of the second element, $i$, indicates that this is the $i^{th}$
particle lost. This information is used to produce a contiguous particle loss
map which contains indexing information on the lost particles. A final step
uses the particle loss map to swap particles to the end of the particle array,
and the particle count is decremented by number of lost particles. This final
step is launched with a different CUDA thread block decomposition that accounts
for the sparsity of operations. Although the final two steps of this algorithm
contain uncoalesced reads and writes, it is still more efficient than a
straightforward sort-by-key algorithm given the sparsity of operations.

}
\section{Performance of accelerated elements}
\label{sec:performance}
{\parindent 16pt

\begin{table}
  \begin{center}
  \begin{small}
    \begin{tabular}{llllll llllll} \\[4mm] \hline \\[0.5mm]
    \textbf{function} & \textbf{elements} & \pbox{10cm}{\textbf{V100}\\time}
                    & \pbox{10cm}{\textbf{K20c}\\time} 
                    & \pbox{10cm}{\textbf{CPU}\\time} 
		    & \pbox{10cm}{\textbf{V100}\\Accel} 
		    & \pbox{10cm}{\textbf{K20c}\\Accel} 
                    &  \textbf{Hist} & \textbf{Conv} & \textbf{Redu} 
                    & \textbf{Loss} & \textbf{Rand} \\[4mm] \hline \\[0.5mm]
      accumulate\_beam\_sums    & main loop     & 4.08 & 24.8  & 696   & 170 & 28.1 
                                              & & & X & & \\[2mm]
    addCorrectorRadiationKick & \pbox{10cm}{HKICK, VKICK,\\KICKER}
					      & 0.009 & 0.066 & 3.97  & 441 & 60.4
                                              & & & X & & X \\[2mm]
      beam\_scraper             & SCRAPER       & 40.9 & 362 & 5249 & 128 & 14.5 
                                              & & & & X & X \\[2mm]
      center\_beam              & CENTER        & 0.41 & 2.13  & 94.8  & 231 & 44.5
                                              & & & X & & \\[2mm]
      elliptical\_collimator    & ECOL          & 0.31 & 2.63  & 78.9  & 257 & 30.0 
                                              & & & X & X & \\[2mm]
      exactDrift                & EDRIFT        & 0.17 & 0.835 & 17.8  & 107 & 21.3
                                              & & & & & \\[2mm]
      limit\_amplitudes         & main loop     & 0.004& 0.022 & 1.14  & 285 & 50.9 
                                              & & & X & X & \\[2mm]
    multipole\_tracking2      & \pbox{10cm}{KOCT, KQUAD,\\KQUSE, KEXT}
					      & 0.28 & 3.33  & 148   & 531 & 44.3
                                              & & & X & X & X \\[2mm]
      rectangular\_collimator   & RCOL          & 0.27 & 1.23  & 59.3  & 220 & 48.1
                                              & & & X & X & \\[2mm]
      simple\_rf\_cavity        & RFCA          & 0.75 & 3.51  & 164   & 218 & 46.6 
                                              & & & X & & \\[2mm]
    track\_particles (M1)     & \pbox{10cm}{DRIFT, QUAD,\\SOLE} 
					      & 0.42 & 1.98  & 69.9  & 165 & 35.3
                                              & & & & & \\[2mm]
    track\_particles (M2)     & \pbox{10cm}{DRIFT, QUAD,\\SOLE}
					      & 0.015 & 0.087 & 13.3  & 887 & 152
                                              & & & & & \\[2mm]
      track\_particles (M3)     & QUAD, SOLE    & 4.46 & 33.2  & 3450  & 774 & 104
                                              & & & & & \\[2mm]
      track\_through\_csbend    & CSBEND        & 1.84 & 53.4  & 1630  & 885 & 30.5 
                                              & & & X & X & X \\[2mm]
      track\_through\_csbendCSR & CSRCSBEND   & 29.7 & 406   & 22k & 742 & 53.7
                                              & X & & X & X & X \\[2mm]
      track\_through\_driftCSR  & CSRDRIFT & 1.26 & 5.25  & 125 & 99.3 & 23.7 
                                              & X & & X & & \\[2mm]
      track\_through\_lscdrift  & LCSDRIFT      & 0.48 & 3.28  & 145 & 302  & 44.1
                                              & X & & X & & \\[2mm]
      track\_through\_matter    & MATTER        & 38.0 & 162   & 3716  & 97.7 & 23.0
                                              & & & & X & X \\[2mm]
      track\_through\_rfcw      & RFCW          & 46.9 & 181   & 7159  & 153 & 39.6
                                              & X & X & X & & \\[2mm]
      track\_through\_trwake    & TRWAKE        & 1.06 & 2.36  & 37.1  & 35.0 & 15.7
                                              & X & X & X & & \\[2mm]
      track\_through\_wake      & WAKE          & 0.34 & 1.33  & 25.6  & 74.4 & 19.2
                                              & X & X & X & & \\[2mm]
    \end{tabular}
  \caption{Elegant kernel performance with 3.2 million particles on NVIDIA
    K20c (CUDA 5.5) and V100 (CUDA 9.0) GPUs and a Intel Core i7-3770K CPU, both using double
    precision. The element column indicates the element name(s) in beamline
    lattice input file that will trigger a call of the kernel. The M\#
    suffix on the \textit{track\_particles} functions indicates the order of
    the matrix.  Times are reported in milliseconds. The last five columns
    list components of the framework used within each function: histograms,
    array convolutions, reductions, particle losses and random number
    generation.}
  \label{tab:kernelPerf}
  \end{small}
  \end{center}
\end{table}

Table~\ref{tab:kernelPerf} lists typical timings and speedup factors (columns
3-7) of accelerated Elegant functions (column 1) and the associated beamline
element name(s) as used in the Elegant lattice input file (column 2)~\cite{ElegantManual}. 
A portion of the accelerated functions are not associated with one particular
element, but rather are called from Elegant's main \textit{do\_tracking} loop.
It is a requirement that these functions are accelerated in order to avoid a
transfer of the particles between the host and the device and thus to avoid an
associated slow-down from effects related to Amdahl's argument
\cite{amdahl67}. For example, consider the \textit{accumulate\_beam\_sums}
function, which computes properties of the beam distribution through a series
of reductions. Depending on the input parameters, these properties may be
computed after each element in the beam line which would be detrimental to an
accelerated computation if this function were not accelerated.

Other beamline elements listed in Tab.~\ref{tab:kernelPerf} are the HKICK and
VKICK, a horizontal and vertical steering dipole elements implemented as a matrix, up to
$2^{nd}$ order; KICKER, a combined horizontal-vertical steering magnet
implemented as a matrix, up to $2^{nd}$ order; SCRAPER, a collimating element
that sticks a limiter into one side of the  beam; CENTER, an element that centers
the beam; RCOL and ECOL, rectangular and elliptical collimator elements;
EDRIFT, an exact drift element; KQUAD, KSEXT, KOCT, and KQUSE, a canonical kick
quadrupole, sextupole, octupole, elements and an element combining quadrupole
and sextupole fields, using either $2^{nd}$ or $4^{th}$  order symplectic
integration; QUAD and DRIFT, quadrupole and drift elements, implemented as a
transport matrix, up to $3^{rd}$ and $2^{nd}$ order, respectively; SOLE, a
solenoid element; CSBEND, a canonical kick sector dipole; CSRCSBEND, a
canonical kick sector dipole with coherent synchrotron radiation;  CSRDRIFT, a
follow-on element to CSRCSBEND that applies the coherent synchrotron radiation
wake over a drift; LSCDRIFT, longitudinal space charge impedance element;
MATTER, a Coulomb-scattering and energy-absorbing element simulating material
in the beam path; the longitudinal and transverse wake elements WAKE and
TRWAKE; the first-order-matrix RF cavity with exact phase dependence RFCA; and
the RF cavity element RFCW, which is a combination of RFCA, WAKE, TRWAKE, and
LSCDRIFT. 

In general, the V100 performance listed in Tab.~\ref{tab:kernelPerf} is greater
than four times better than the K20c.  This is consistent with the factor of
four greater flop rate and enhanced memory bandwidth for the V100 relative to
the K20c.  The V100 is the next generation GPU to be included in new Oak Ridge
supercomputer, Summit, while the K20c is roughly comparable to the K20x GPUs
available on the Titan Cray XK-7 supercomputer also at Oak Ridge.

\begin{figure}
\centering
  \includegraphics[width=0.45\linewidth]{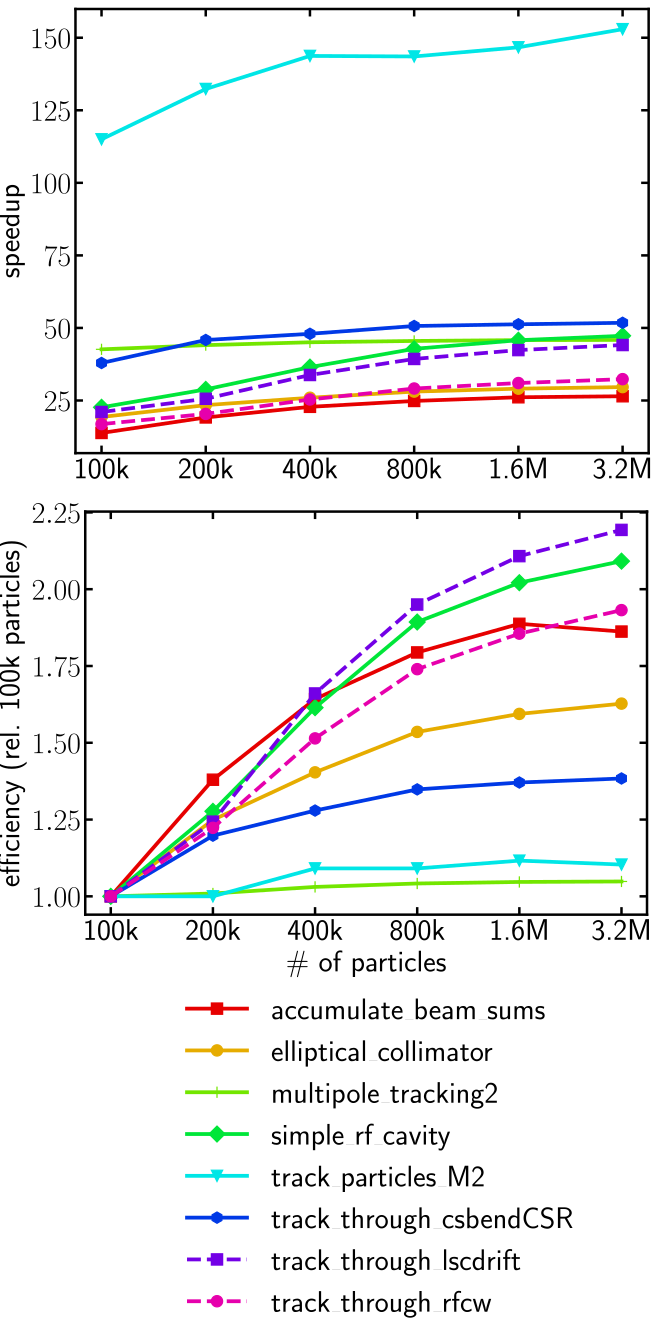}
  \caption{Scaling of speed-up factors (top panel) and efficiency (assuming $O$(number of
    particles) scaling, bottom panel) relative to the number of particles in the computations for
    selected kernels. Comparisons are for a single NVIDIA Tesla K20c GPU versus
    an Intel Core i7-3770K CPU.}
  \label{fig:partScale}
\end{figure}

Best performance, relative to the CPU code, is achieved with a large number of
particles (hundreds of thousands 
and up). The computational cost of most kernels
scales as $O(N)$ where $N$ is the number of particles. When $N$ is large, other
constant-scaling components (e.g. kernel launch overhead and array
convolutions) are amortized over the large $O(N)$ number of computations and have little
impact on the GPU performance. Fig.~\ref{fig:partScale} plots the speed-up
factors and efficiency scaling relative to the number of particles in the
computations for selected accelerated functions.  In general, good speed-up
factors of 15-150x are obtained for these functions over a broad range of
particles (100k-3.2M).  The efficiency is measured relative to the case with
100k particles assuming constant-in-N scaling. As the number of particles is
increased, the efficiency curves decrease in slope, indicating the increased
amortization of the constant scaling components begins to have little effect.
The \textit{multipole\_tracking2} and \textit{track\_particles\_M2} functions do
not use either histograms or reductions thus have a very small slope compared
to the other functions that either use reductions or incorporate collective
effects through histograms and convolutions. 

}
\section{Distributed-memory scaling}
\label{sec:scaling}
{\parindent 16pt

Next we describe performance and scaling studies of the GPU-accelerated Elegant
relative to the CPU-only version of the code. These studies are performed on
the 18,688-node, hybrid-architecture, Titan Cray XK-7 supercomputer at the Oak
Ridge Leadership Computing Facility at the Oak Ridge National Laboratory.  We
use the Linac Coherent Light Source (LCLS)~\cite{LCLShome} beam delivery
system~\cite{LCLS_CDR_Ch7, LCLSlattice} as our test lattice, such that the
studies here represent end-to-end application performance in a realistic
setting, as opposed to the kernel and function-specific descriptions from
previous sections.  In order to have a more balanced comparison of the GPU to
CPU performance, the performance of two 8-core AMD Opteron CPUs is compared to
the performance of a single NVIDIA Tesla K20x (as there are 16 CPU cores and a
single GPU per Titan node). Both the CPU-only and GPU-enabled versions of
the code are compiled with SDDS parallel (MPI) I/O enabled.

\begin{figure}
\centering
  \includegraphics[width=0.5\linewidth]{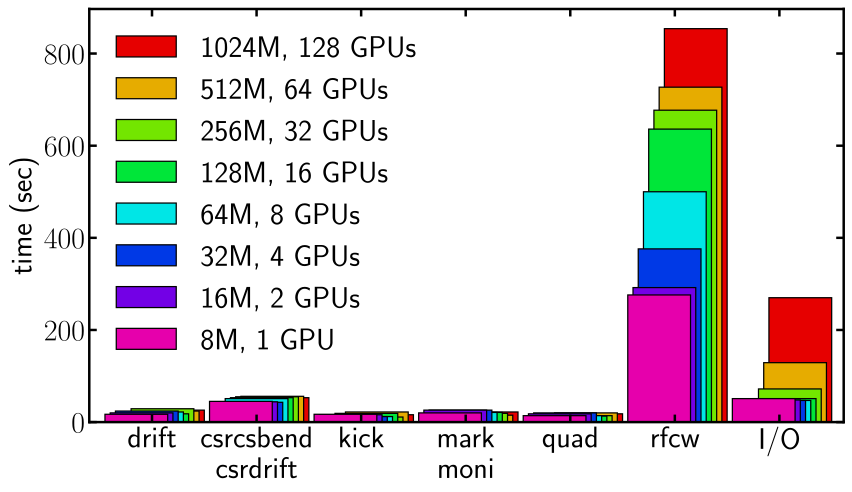}
  \caption{Weak scaling results for the GPU-accelerated version of Elegant,
    arranged by beamline element. LCLS driver linac lattice is used as the test
    case for this study and 8 million particles are used per GPU.}
  \label{fig:titanGpuWeak}
\end{figure}

Results of the weak scaling studies (where the number of cores is increased in
proportion to the problem size) are shown in Fig.~\ref{fig:titanGpuWeak}. One
can see that most beamline elements exhibit nearly perfect scaling over the
explored range of the problem sizes (up to one billion macroparticles). The
exception is the RFCW kernel, which is limited by the parallel scaling of its
component transverse-wake kernel. One observation from these weak scaling
studies is that, for the full LCLS test case, and relative to a job with 8
million particles, it takes the GPU-accelerated code only two and a half times
longer to run a $256\times$ bigger job to completion, a very good scaling
performance in this important-in-practice range of problem sizes. In particular,
the full LCLS beamline simulation was done within 18.5 minutes when
using one billion ($10^9$) particles, which is comparable to the number 
of actual electrons in the beam.

\begin{figure}
\centering
  \includegraphics[width=0.5\linewidth]{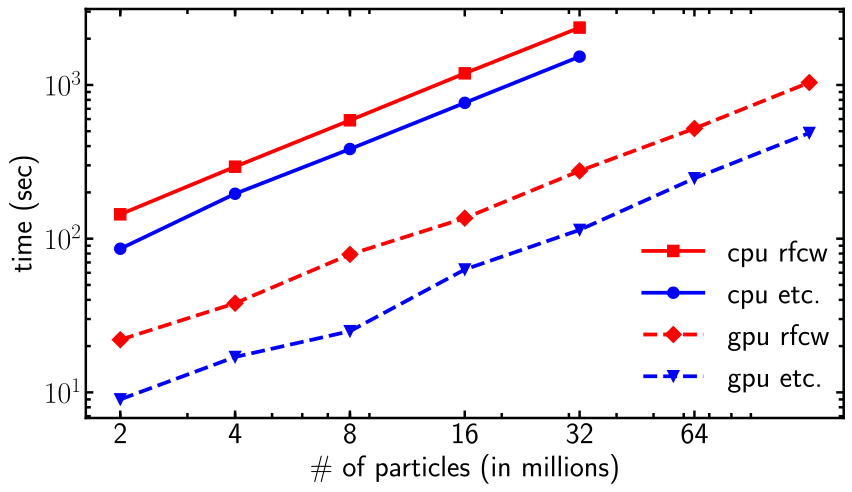}
  \caption{Scaling with the number of simulation particles for small to
    moderate particle counts on a single node of Titan for equivalent GPU (dashed
    lines) and CPU (16 cores, solid lines) runs. The RFCW kernel is plotted
    independently as it dominates the run time, all other elements are included in
    the etc.~lines. Timings show I/O takes an order of magnitude less time than the
    plotted lines. The LCLS beam delivery system's lattice is used as a test case.}
  \label{fig:titanParticle}
\end{figure}

Figure \ref{fig:titanParticle} shows the results of increasing the number of
particles while running on only a single Titan node. The RFCW kernel is plotted
independently as it dominates the run time, and all other elements are grouped
as a single line (labeled ``etc.''). The CPU-only (16 cores) and GPU-accelerated
versions of the code are run for small-to-moderate particle counts. The scaling
is linear as expected for ideal scaling and the GPU-accelerated version
consistently outperforms the CPU-only version in terms of time to solution for
a start-to-end simulation. For example, with 8 million particles the CPU-only
version requires 64 minutes, whereas the GPU-accelerated version runs in 7
minutes, a speed-up of nearly a factor of 10.

\begin{figure}
\centering
  \includegraphics[width=0.5\linewidth]{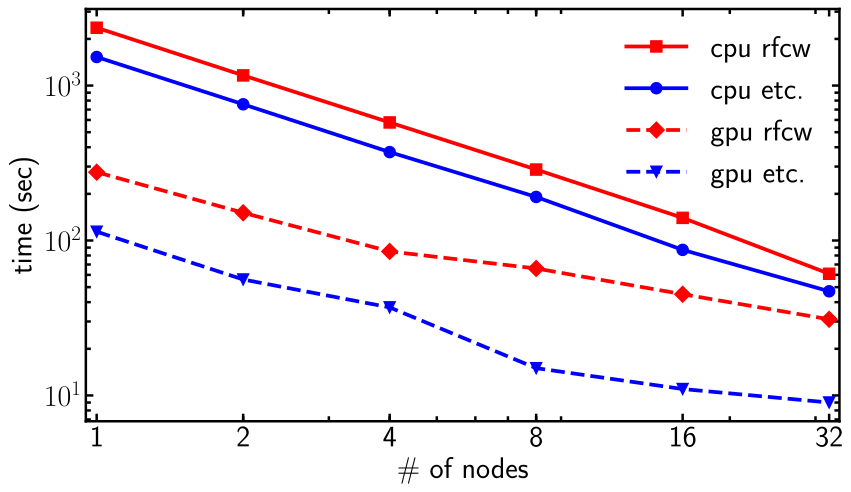}
  \caption{Strong scaling results for 8 million particles on Titan for
equivalent GPU (1 GPU/node, dashed lines) and CPU (16 cores/node, solid lines)
runs. The LCLS driver linac's lattice is used as a test case. }
  \label{fig:titanCpuStrong8M}
\end{figure}

\begin{figure}
\centering
  \includegraphics[width=0.5\linewidth]{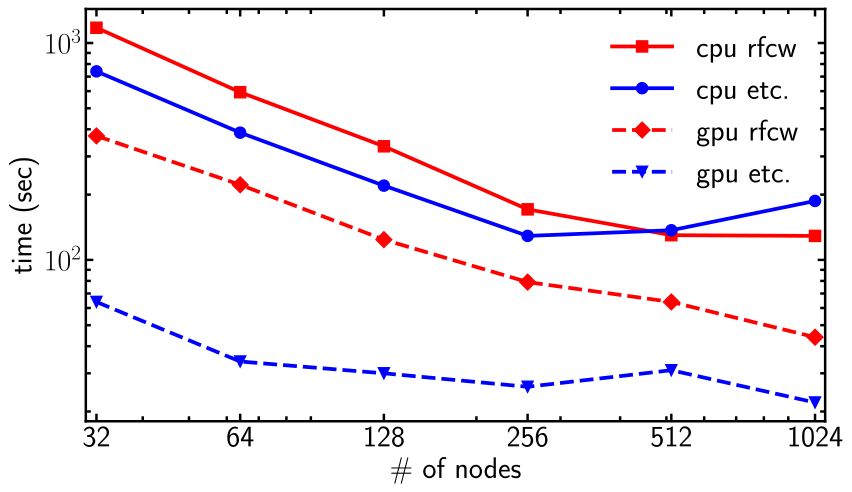}
  \caption{Strong scaling results for 128 million particles on Titan for
equivalent GPU (1 GPU/node, dashed lines) and CPU (16 cores/node, solid lines)
runs. Performance is not significantly improved for job sizes greater than
256 nodes. The LCLS beam delivery system's lattice is used as a test case.}
  \label{fig:titanCpuStrong128M}
\end{figure}

Next, we present strong scaling studies (increasing the number of cores for a
fixed problem size), once again using the LCLS driver linac's lattice as a test
case.  Figures \ref{fig:titanCpuStrong8M} and \ref{fig:titanCpuStrong128M} show
the scaling of total time spent in computationally expensive RFCW and in the
rest of the elements, for 8 million particle runs
(Fig.~\ref{fig:titanCpuStrong8M}) and 128 million particle runs
(Fig.~\ref{fig:titanCpuStrong128M}).  In general, the GPU version outperforms
the CPU version at all job sizes.  Both the CPU and GPU versions exhibit good
scaling with 8 million particles as shown in Fig.~\ref{fig:titanCpuStrong8M}.
For the 128 million particle case showing in Fig.~\ref{fig:titanCpuStrong128M},
there is a scaling bottleneck at approximately 256 nodes (~4000 cores) with the
CPU version of the code. The equivalent scaling of the GPU enabled code is
degraded but still beneficial at these large job sizes.

As regards comparing the performance of the GPU-enabled Elegant to that of the
parallel CPU-only version, one useful metric may be to compare the scaling of
the number of cores needed for the two versions of the code to achieve
approximately the same time to solution, as the problem size varies.  For good
performance by the accelerated code, a sufficiently large number of particles
is required on the GPU. Thus, this metric is most relevant when large particle
counts are necessary. We found that, for a 1M-particle simulation of the LCLS
lattice, a simulation on 1 GPU takes the same time as a simulation on 16 CPUs,
and an 8M-particle run on a single GPU takes roughly the same time as a 100-CPU
run. Timing on Titan with 128M particles resulted in 16 K20 Kepler GPUs being
equivalent to 1024 16-core AMD Opteron CPUs (again, in terms of time to
solution).  Thus, one could argue that, by this metric, a GPU cluster
would be a more cost-efficient hardware choice for this type of simulation.

}
\section{Code verification}
\label{sec:verification}
{\parindent 16pt

Testing and verification of the CUDA implementation to ensure accuracy of the
results has been an essential part of the work reported here.  This is a
complicated task, given that the potential Elegant use cases are widely varied,
and full coverage of all cases is difficult. We describe four
methods that are used to verify the GPU code: memory synchronization, unit
tests, run-time CPU/GPU particle phase space coordinate comparisons after each
element, and end-to-end distribution properties comparisons. We recommend that
users of GPU Elegant employ these methods as appropriate  to test new beamline
setups with a modest number of particles before performing production runs.

\subsection{Memory synchronization}

A general consideration of any GPU/CPU implementation is synchronization and
access limitations of data between the host and device memory. A logic error
where both the CPU and GPU access and modify the particle data without
synchronization would clearly have a detrimental effect on the simulation
accuracy.  In the worst case scenario, this logic error can lead to a
situation where the application runs and produces a reasonable-looking but
incorrect result. One mechanism for handling this problem is to accept that a
run-time segmentation fault is better than producing the wrong result.  Thus,
when the particles are present on the device memory, the host particle array
pointer is nullified. If any host operation dereferences the particle array
pointer a segmentation fault is triggered. This provides the user with an
indication that the result cannot be trusted, and gives the developer
detailed information on the location of the logic error.

\subsection{Unit tests} \label{sec:unit}

Unit tests are designed to test a specific component of the code, have
well-defined input and output, and run quickly. Ideally, unit tests should have
full code coverage in the sense that if a mistake is introduced anywhere in the
code during development, one or a set of the unit test should fail.  It is also
desirable for the unit test to have a hierarchical order such that the first
unit test to fail should immediately point to the underlying problem that may
cause the subsequent tests to fail. Our tests do not live up to this idealized
standard, but instead are limited in scope to testing the GPU infrastructure
described in Sec.~\ref{sec:infrastructure}. These tests initialize a particle
distribution, or general arrays of data and run the reduction, histogram,
convolution, and particle-loss algorithms on both the CPU and GPU. A test is
considered a failure if either the resulting output does not match or the GPU
fails to outperform a single CPU core.

\subsection{Run-time host/device element comparisons} \label{sec:runtime}

In order to verify the code with a wide variety of beamline element use cases,
runtime verification of the GPU routines can be enabled via a preprocessor flag
(GPU\_VERIFY=1) during compilation. Compilation with GPU verification is not
intended for production runs, but rather it is to be used as an aid in
development and as an option for the user to check the code.

With respect to programming details, the GPU function calls are embedded within
the equivalent CPU routines, such that the GPU version is used when Elegant is
compiled with GPU acceleration. When verification is enabled, timings are
computed with the CUDA event timers where, in addition to the CUDA version of
the routine, the CPU routine is also recursively called and timed. An example
of the GPU and verification hooks is shown in
Code~\ref{code:cpuExactDriftHooks}.  The result from both routines is then
copied to the host memory and compared for accuracy. Warnings are printed if
the resulting particle phase-space coordinates do not agree within a tolerance
of $10^{-10}$. At the end of the Elegant run, aggregate timing statistics are
displayed for each accelerated routine called.

\begin{lstlisting}[frame=single, language=c++,
  caption={GPU and verification callback hooks within the CPU exactDrift code.},
  label={code:cpuExactDriftHooks}]
#ifdef HAVE_GPU
  if(getElementOnGpu()){
    startGpuTimer();
    gpu_exactDrift(np, length);
#ifdef GPU_VERIFY     
    startCpuTimer();
    exactDrift(part, np, length);
    compareGpuCpu(np, "exactDrift");
#endif /* GPU_VERIFY */
    return;
  }
#endif /* HAVE_GPU */
\end{lstlisting}

This run-time testing has several advantages and disadvantages relative to unit
tests. Unit tests can be quickly run just after compilation and provide simple
test cases for debugging, both advantages over the run-time testing. However,
run-time testing allows for full coverage of possible use cases and can be more
easily integrated within the existing Elegant regression testing system than
unit tests. The implementation of the run-time testing is vastly more
straightforward than unit testing, as all the element information is already
present and does not need to be configured by the testing framework. One
drawback to run-time tests is that some algorithms can generate significant
false positives. For example, consider an algorithm used to center the beam at
location $\mathbf{x}_c$ via the operation $\mathbf{x}_i = \mathbf{x}_i -
\mathbf{x}_c$ for all particles, $i$, at locations $\mathbf{x}_i$. For a large
number of particles, some fraction of particles may already be almost at the centering
location, $\mathbf{x}_c$, and thus there will be a near-exact cancellation
resulting in most of the floating-point significant digits to be filled with
round-off values. This round-off error almost certainly does not agree during
comparisons between the CPU and the GPU, and can produce spurious errors of
order unity.

\subsection{End-to-end verification} \label{sec:end2end}

The final quality assurance test that we employ focuses on statistical
properties of the particle beam distribution in start-to-end runs.  For
statistical tests to be meaningful, they should be applied either to the
whole beam or to beam slices that constitute a large sample in terms of the
macroparticle count (say, 50,000 particles or more).  These tests are
essential for kernels that use random numbers, as well as in simulation
settings where accumulation of round-off is a concern. 

There are a number of elements in Elegant that employ a random number generator
(RNG) for simulation of one or more physical effects.  For example, a
Coulomb-scattering and energy-absorbing element MATTER that simulates material
in the beam path uses an RNG for computing the probability of scattering for
each particle and the scattering angle, and the CSRCSBEND and CSBEND elements'
incoherent synchrotron radiation model uses a random number to determine the
number of photons emitted and then requires an additional random number for
each emitted photon.  For the GPU version of the code, we implement a random
number generation framework which is based on the CUDA cuRAND library.  Just as
in the host CPU code, the random-number sequence generated in the GPU code is
reproducible if the same seed is used in a subsequent run.  However, the random
number sequences themselves are not the same in the CPU and GPU code, which
automatically causes particle-by-particle comparison tests to fail.

Statistical tests are facilitated by the simulation output data being in the
SDDS format, so that one can make use of of several tools from the SDDS Toolkit
for extracting, sorting, analyzing and visualizing the particle data.  In
addition, statistical tests can employ a variety of Elegant's built-in
capabilities for generating runtime beam phase-space distribution
statistics. 
}
\section{Summary and Conclusion}
\label{sec:conclusion}
{\parindent 16pt

We developed a GPU-accelerated version of particle accelerator code Elegant.
The new version demonstrates a greatly improved performance on hybrid
platforms for computationally intensive simulations of the kind that
frequently arises in the course of design and optimization of particle
accelerator-based systems.  A prominent feature of the computational
infrastructure of the GPU-enabled Elegant is a C++ templated class framework.
This framework facilitates the creation of extensible kernels and provides
abstract interfaces that simplify the implementation of particle operations
(e.g., by hiding the conversion between the array-of-structures and
structure-of-arrays formats of the particle data).  

We implemented optimized kernels for the reductions, convolution, histogram
computation, and other operations that are at the core of modeling collective
effects in Elegant and whose suboptimal performance can be the main limiting
factor to the overall performance of the code.  Our optimized histogram kernel
creates sub-histograms in shared memory in such a way as to reduce the thread
contention while maintaining high block occupancy, combining the sub-histograms
from different blocks to produce the final histogram.  The convolution
computation kernel (required, e.g., in the computation of wakefield effects)
relies on buffering sub-sections of each array in shared memory for computing
part of the result for a given array index, with each thread block applying an
atomic addition to produce the final result for the convolution.  Reduction
operations are central to the computation of the statistical properties of
particle distributions.  In our implementation, standard reduction algorithms
are templated over the reduction operation.  Our framework allows for launching
asynchronously on separate CUDA streams of functions that reduce quantities
from separate data arrays.  We find asynchronous reductions to be about 40\%
faster than their synchronous counterparts on Tesla K20c GPUs with 1M-particle
distributions.  Finally, the modeling of beam particle loss in traversing the
beamline necessitated a complete re-working for the GPU version of the
CPU-based algorithm so as to accommodate concurrent particle-update operations.
A specialized particle loss sorting algorithm described in this paper was
optimized for the physically relevant case where the fraction of lost particles
is small, and in our tests it was significantly faster than a straightforward
sort-by-key algorithm, the ratio depending on the fraction of particles lost
(e.g., 4$\times$ faster for the case of 0.5\% particle loss). 

The performance benefit from porting to the GPUs is, of course, only realized
when the number of simulation particles is sufficiently large.  In the case of
GPU-accelerated Elegant, we find that the cost of kernel launches and
constant-scaling components such as convolutions is well amortized for a
particle count that is element-dependent, but generally in the range of 100k to
about 3.2M.  This range is well below the number of particles used
in large-scale simulations with Elegant, and and it should be kept in mind when
allocating resources to a simulation with the GPU version of the code.
Regarding the scaling behavior, in tests with up to one billion particles we see
essentially ideal weak scaling for all elements except RFCW (which exhibited
good scaling nonetheless, running only 2.5$\times$ slower for a
256$\times$ larger problem in what currently is a very common range of
problem sizes).  

From the perspective of the end user, of primary interest is not so much the
speed-up data for individual kernels, but the performance of the code in
start-to-end (S2E) simulations in realistic settings.  We compared the
performance of the GPU-enabled and CPU-only versions of Elegant by running S2E
simulations of the LCLS beam delivery linac (fairly typical of an important
class of accelerator systems) on Titan Cray XK-7 at ORNL.  Focusing on
the scaling of the two versions of the code in terms of the number of cores
needed to achieve the same time to solution in S2E simulations, we found the
GPU version to increasingly outperform the CPU-only version as the problem size
grows larger: For example, performing a 1M-particle simulation of the LCLS
lattice on Titan, a run on 1 K20 Kepler GPU takes as long as a run on 16
16-core AMD Opteron CPUs, while a 128M-particle run requires either 16 GPUs or
1024 CPUs to achieve the same time to solution.

In response to the evolving needs of the accelerator physics research
community, new simulation capabilities are continually added to Elegant over
time.  To facilitate the maintainability and continued development of the code,
the GPU version's source code is organized in such a way that it can be
easily related to the corresponding code in the CPU implementation.  For
quality assurance purposes we developed a testing and verification
infrastructure that includes unit tests, runtime GPU/CPU phase space
coordinate comparisons after individual elements, comparisons of the
particle distribution statistical properties in end-to-end runs, and a
mechanism for handling the memory synchronization problems.  We made a
heavy use of this testing and verification framework in the process of
developing the GPU version of the code, and we expect it to be of value as
new capabilities are added to Elegant in the future. 

}

\section{Acknowledgements}
\label{sec:acknowledgements}
{\parindent 16pt

This work was supported by the US DOE Office of Science, Office of Basic Energy
Sciences under grant number DE-SC0004585, and in part by Tech-X Corporation,
Boulder, CO. This research used resources of the Oak Ridge Leadership Computing
Facility at the Oak Ridge National Laboratory, which is supported by the
Office of Science of the U.S. Department of Energy under Contract No.
DE-AC05-00OR22725.  We would like to thank James Balasalle and Chris DeLuca
for assistance with development.  

}

\bibliographystyle{elsarticle-num}   
\bibliography{biblio}

\end{document}